\documentclass[prd,showpacs,amsmath,amssymb]{revtex4}

\usepackage{bm}

\begin{document}

\title{Inflation and accelerated expansion TeVeS cosmological solutions}

\author{Luz Maria Diaz-Rivera, Lado Samushia, and Bharat Ratra}

\affiliation{Department of Physics, Kansas State University, 116
Cardwell Hall, Manhattan, KS 66506, USA}

\date{January 2006 \;\;\;\;         KSUPT-06/1}

\begin{abstract}

We find exact exponentially expanding and contracting de Sitter solutions 
of the spatially homogeneous TeVeS cosmological equations of motion in the 
vacuum TeVeS model and a power law accelerated expanding solution in the presence 
of an additional ideal fluid with equation of state parameter $-5/3 < \omega < -1$. 
A preliminary stability analysis shows that the expanding vacuum
solution is stable, while in the ideal fluid case stability
depends on model parameter values. These solutions might provide a
basis for incorporating early-time inflation or late-time
accelerated expansion in TeVeS cosmology.
\end{abstract}


\maketitle

\section{Introduction}
\label{sec:intro}

For more than seven decades hypothetical dark matter has been
postulated to explain observed large-scale velocities that are
larger than expected on the basis of Newton's second law of
motion, Newton's inverse-square law of gravitation, and the
assumption that  cosmological structures under investigation are
in gravitational equilibrium, \cite{zwickly, trimble, peebles1}.
Great effort has been devoted to the search for direct evidence of
dark matter, \cite{bertone}, and  although there have been
advances there is no conclusive direct evidence for dark matter.

Observations disfavor the alternate possibility that the
gravitational force falls off slower with distance than Newton's
inverse-square law predicts and that there is not a significant
amount of dark matter (see Secs. IV.A.1 and IV.B.13 of Ref.
\cite{peebles}).

Another explanation --- modified Newtonian dynamics (MOND) --- has
been proposed by Milgrom \cite{milgrom65}. In MOND, Newton's
second law, $\vec a = -  \nabla \Phi_N$, is modified to

\begin{equation}
\label{milgrom}
\mu(| {\vec a}|/a_0)\vec a = -\nabla \Phi_N.
\end{equation}

\noindent Here $\vec{a}$ is acceleration, $a_0$ is an acceleration
scale, $\Phi_N$ is the Newtonian potential, and $\mu(x)$ is a
function which satisfies $\mu(x) \approx 1$ when $x \gg 1$, and
$\mu(x) \approx x$ when $x \ll 1$. From large-scale galactic data
it has been estimated that $a_0 \approx 1 \, \times  \, 10^{-8}$
${\rm cm \, s^{-2}}$, so in the solar system where accelerations
are large compared to $a_0$ the usual Newtonian law is recovered.
Equation (\ref{milgrom}) and this choice of $\mu(x)$ was proposed
to explain the observed flat rotation curves of disk galaxies,
without need for dark matter. MOND also explains the Tully-Fisher
relation, the observed correlation between the infrared luminosity
of a disk galaxy and the rotation velocity in the flat region of
the rotation curve. See Refs. \cite{bek-mil} for other tests and
predictions of MOND.

In spite of its success, MOND is not a consistent theory and has
some apparent problems. MOND requires dark matter in galaxy
clusters, \cite{sanders03}, and if this is massive neutrinos the
needed neutrino mass is on the edge of being ruled out
\cite{elgaroy}. There are indications that observations require
that the value of $a_0$ depend on the object being studied
\cite{famaey}. Other possible problems are considered in Refs.
\cite{zhao05}. A major drawback of MOND was the lack of a
relativistic generalization needed for a consistent MOND
cosmological model.

Bekenstein \cite{bekenstein} has recently proposed a new covariant
field theory  which has MOND characteristics in the weak
acceleration limit and provides a setting for constructing
consistent cosmological models. This model uses three dynamical
gravitational fields --- a tensor, a vector, and a scalar field
--- leading to the acronym TeVeS. Several authors have studied
different aspects  of the TeVeS model, including large-scale
structure formation \cite{skordis}, gravitational lensing
\cite{chiu}, and the strong gravity regime of the model
\cite{giannios}. Hao and Akhoury \cite{hao} have studied a
cosmological model and its dynamics in the context of TeVeS , but
no exact solution has been found so far.

Whether TeVeS or some generalization is able to provide an
accurate model of cosmology is an issue of some current interest.
The standard dark energy and dark matter dominated general
relativistic cosmological model does a remarkable job on large
scales, however, it appears to have trouble fitting some smaller
scale observations \cite{peebles}. TeVeS cosmology on the other
hand is only now attracting preliminary attention. In the standard
cosmology the Universe goes through two periods of accelerated
expansion, at early times during inflation and during the current
epoch. Inflation is used to make the Universe homogeneous and to
generate the quantum-mechanical fluctuations responsible for
observed large-scale structures \cite{hawking} while supernova and
other data indicate that the expansion is speeding up now
\cite{peebles}. It is of interest to understand whether TeVeS
cosmology has an accelerated solution that might play a similar
role to that in the standard model. A complete and consistent
TeVeS cosmological model would be very useful, at the very least
providing stimulus for the development of precision tests that can
be used to distinguish the TeVeS predictions from those of
standard cosmology, along the lines of observational tests that
were developed to distinguish open inflation from the standard
flat model \cite{gott}.

The paper is organized as follows. In Sec. \ref{sec:equations} we
present the general TeVeS equations of motion and in Sec.
\ref{sec:cosmology} we specialize these equations to the spatially
homogeneous and isotropic Friedmann-Lema\^itre-Robertson-Walker
spacetime model. In Sec. \ref{sec:sol} we present exact solutions
of these equations and discuss cosmological implications. We
examine the stability of these solutions in Sec.
\ref{sec:stability}. Finally, in Sec. \ref{sec:conclusion} we
summarize our results and present conclusions.

\section{TeVeS equations of motion}
\label{sec:equations}

TeVeS assumes three dynamical gravitational fields: a tensor
metric $g_{\mu \nu}$, a unit norm time-like four-vector
$U_{\alpha}$, and a scalar $\phi$. There is also a nondynamical
scalar field $\sigma$. The three dynamical fields are connected
through the physical metric tensor

\begin{equation}
\label{eq:phys-met}
\tilde{g}_{\alpha \beta}= e^{-2\phi}g_{\alpha
\beta}-2U_{\alpha}U_{\beta}\sinh(2 \phi) = e^{-2\phi}(g_{\alpha
\beta}+ U_{\alpha}U_{\beta})- e^{2\phi}U_{\alpha}U_{\beta},
\end{equation}

\noindent where $g_{\alpha \beta}$ is the Einstein metric tensor.
Our conventions are those of Bekenstein \cite{bekenstein} with
metric signature diag$(-1,1,1,1)$.

Following Bekenstein the action of the system is written as

\begin{equation}
\label{eq:action}
S=S_g+S_s+S_v+S_m,
\end{equation}

\noindent where the actions for the tensor field $S_g$, the two
scalar fields $S_s$, the vector field $S_v$, and ordinary matter
(as opposed to Bekenstein's $\phi$, $\sigma$, and $U_\alpha$
fields) $S_m$, are,

\begin{eqnarray}
\label{eq:action-g}
S_g &=& \frac{1}{16\pi G}\int{d^{4}\!x \sqrt{-g} \, g^{\alpha \beta}R_{\alpha \beta}}, \\
\label{eq:action-s}
S_s &=& -\frac{1}{2}\int{d^4\!x  \sqrt{-g}
\left[\sigma^2 h^{\alpha \beta}\phi_{,\alpha}\phi_{,\beta}+\frac{G}{2 l^2}\sigma^4 F(kG\sigma^2)\right]}, \\
\label{eq:action-v}
S_v &=& -\frac{K}{32 \pi G}\int{d^4\!x \sqrt{-g}
\left[g^{\alpha \beta}g^{\mu \nu} U_{[\alpha,\mu]}U_{[\beta,\nu]}-
\frac{2 \lambda}{K}\left(g^{\mu \nu}U_{\mu}U_{\nu}+1\right)\right]}, \\
\label{eq:action-m}
S_m &=& \int{d^4\!x  \sqrt{-\tilde{g}} \,
{L(\tilde{g}_{\mu \nu}, f^{\alpha}, f^{\alpha}_{|\mu}, \dots)}}.
\end{eqnarray}

\noindent Here $g$ and $R_{\alpha \beta}$ are the determinant and
Ricci tensor constructed from the Einstein metric tensor
$g_{\alpha \beta}$, the tensor $h^{\alpha \beta} = g^{\alpha
\beta}-U^{\alpha} U^{\beta}$ has been used instead of $g^{\alpha
\beta}$ in order to avoid acausal dynamical scalar field
propagation \cite{bekenstein2}, and $f^{\alpha}$ represents
ordinary matter, including possibly a scalar field or a
cosmological constant that drives accelerated expansion or
inflation. Also, $F$ is a free dimensionless function, $K$ and $k$
are positive dimensionless coupling constants, $l$ is a constant
scale of length, $\lambda$ is a spacetime dependent Lagrange
multiplier, and $G$ is the bare gravitational constant. Square
brackets in Eq. (\ref{eq:action-v}) denote antisymmetrization,
$A_{[\mu}B_{\nu]}=A_{\mu}B_{\nu}-A_{\nu}B_{\mu}$, in what follows
round brackets denote symmetrization,
$A_{(\mu}B_{\nu)}=A_{\mu}B_{\nu}+A_{\nu}B_{\mu}$, and the
covariant derivative denoted by ``$|$'' in Eq. (\ref{eq:action-m})
is defined in terms of the physical metric tensor.

The equations of motion are obtained by varying the action given
in Eq. (\ref{eq:action}) with respect to $U_{\alpha}$, $\sigma$,
$\phi$, $g_{\alpha \beta}$, and $\lambda$. These equations are,

\begin{equation}
\label{eq:motion-vec}
K(U^{[\alpha ;\beta]}{}_{;\beta}+U^{\alpha}U_{\gamma}U^{[\gamma ;\beta]}{}_{;\beta})+8\pi G
\sigma^2[U^{\beta}{\phi}_{,\beta}g^{\alpha \gamma}\phi_{,\gamma}+U^{\alpha}(U^{\beta} \phi_{,\beta})^2]=8\pi
G(1-e^{-4\phi})[g^{\alpha \mu}U^{\beta}\tilde{T}_{\mu \beta}+U^{\alpha}U^{\beta}U^{\gamma}\tilde{T}_{\gamma \beta}],
\end{equation}

\begin{equation}
\label{eq:motion-s1}
-\mu F(\mu)-\frac{1}{2}\mu^2 F'(\mu)= k l^2 h^{\alpha \beta}\phi_{,\alpha}\phi_{,\beta} ,
\end{equation}

\begin{equation}
\label{eq:motion-s2}
[ \mu (kl^2 h^{\mu \nu}\phi_{,\mu}\phi_{,\nu})h^{\alpha \beta}\phi_{,\alpha}]_{;\beta}
= kG[g^{\alpha \beta}+(1+e^{-4\phi})U^{\alpha}U^{\beta}]\tilde{T}_{\alpha \beta},
\end{equation}

\begin{equation}
\label{eq:motion-g}
G_{\alpha \beta}= 8\pi G[\tilde{T}_{\alpha \beta}+
(1-e^{-4\phi})U^{\mu}\tilde{T}_{\mu (\alpha}U_{\beta)}+\tau_{\alpha \beta}]+ \Theta_{\alpha \beta},
\end{equation}

\begin{equation}
\label{eq:porque?}
g^{\mu \nu}U_{\mu}U_{\nu}=-1.
\end{equation}

\noindent Here $\tilde{T}_{\mu \nu}$ is the stress-energy tensor
of ordinary matter. Equation (\ref{eq:motion-vec}) is the vector
field equation of motion. The prime in Eq. (\ref{eq:motion-s1})
denotes a derivative with respect to $\mu = kG\sigma^2$. Equation
(\ref{eq:motion-s1}) determines $\mu$ and the nondynamical scalar
field $\sigma$ in terms of $g_{\alpha \beta}$,  $U_{\alpha}$, and
$\phi$. Equations (\ref{eq:motion-s2}) and (\ref{eq:motion-g}) are
the equations of motion for the dynamical scalar field $\phi$ and
the metric  tensor respectively. The Einstein tensor $G_{\alpha
\beta}=R_{\alpha \beta}-R g_{\alpha \beta}/2$, where $R$ is the
Ricci scalar constructed from the metric tensor $g_{\alpha
\beta}$. We call the right hand side of Eq. (\ref{eq:motion-g})
$\, \, 8 \pi G \tilde{T}^{\rm{eff}}_{\alpha \beta}$.

The tensors $\tau_{\alpha \beta}$ and $\Theta_{\alpha \beta}$ in
Eq. (\ref{eq:motion-g}) are defined as

\begin{eqnarray}
\label{eq:tau}
\tau_{\alpha \beta} &=&  \sigma^2\left[\phi_{,\alpha}\phi_{,\beta}-\frac{1}{2}g^{\mu \nu}
\phi_{,\mu}\phi_{,\nu}g_{\alpha \beta}-U^{\mu} \phi_{,\mu}\left(U_{(\alpha}\phi_{,\beta)}-
\frac{1}{2}U^{\nu}\phi_{,\nu}g_{\alpha \beta}\right)\right]
 -\frac{G}{4 l^2}\sigma^{4}F(\mu)g_{\alpha \beta}, \\
\label{eq:theta}
\Theta_{\alpha \beta} &=& K\left(g^{\mu \nu}U_{[\mu ,\alpha]}U_{[\nu ,\beta]}-
\frac{1}{4}g^{\sigma \tau}g^{\mu \nu}U_{[\sigma , \mu]}U_{[\tau , \nu]}g_{\alpha \beta}\right).
\end{eqnarray}

\noindent Assuming that ordinary matter is an ideal fluid, the
matter stress-energy tensor in the physical coordinate system is
$\tilde{T}_{\alpha
\beta}=\tilde{\rho}\tilde{u}_{\alpha}\tilde{u}_{\beta}+\tilde{p}(\tilde{g}_{\alpha
\beta}+\tilde{u}_{\alpha}\tilde {u}_{\beta})$, with $\tilde{\rho}$
the energy density, $\tilde{p}$ the pressure, and
$\tilde{u}_{\alpha}$ the fluid four-velocity. An equation of state
$\tilde{p}(\tilde{\rho})$ must be specified to fully determine
this stress-energy tensor. The Bianchi identity $\tilde{T}^{{\rm
eff}\, \, ;\beta}_{\alpha \beta}=0$, together with Eqs.
(\ref{eq:motion-vec})--(\ref{eq:motion-s2}) and
(\ref{eq:porque?}), implies stress-energy conservation for
ordinary matter, which may be used as an equation of motion for
ordinary matter, instead of using part of the tensor equation of
motion, Eq. (\ref{eq:motion-g}). Equation (\ref{eq:porque?}),
which is a constraint equation derived by varying the action with
respect to the Lagrange multiplier $\lambda$, forces $U_{\alpha}$
to be time-like with unit norm.

\section{Spatially homogeneous and isotropic equations of motion}
\label{sec:cosmology}

Observations indicate that on large scales the spatial
distribution of matter and radiation are close to isotropic in the
mean \cite{peebles1}. This fact motivates consideration of the
spatially homogeneous and isotropic
Friedmann-Lema\^itre-Robertson-Walker spacetime model. In general
relativistic cosmology observations also show that space is close
to flat (see, {\it e.g.}, \cite{peebles}). Although it is not yet
clear if observations also favor vanishing spatial curvature in
Bekenstein's model, for simplicity we will consider the
spatially-flat Friedmann-Lema{{\^\i}}tre-Robertson-Walker metric,

\begin{equation}
\label{eq:metric}
d\tilde {s}^2=-d{\tilde{t}}^2+\tilde{a}^2[dr^2+r^2 (d\theta^2+ \sin ^2 \theta \, d\varphi^2)],
\end{equation}

\noindent where $\tilde {dt}=e^{\phi}dt$ and
$\tilde{a}=e^{-\phi}a$ relate the time and scale factor in the
physical and Einstein coordinates. The scalar field $\phi$ depends
on time and since there cannot be a preferred spatial direction
the vector field $U^{\mu}$ must point in the time direction,
$U^{\mu}=\delta^{\mu}_{t}$ \cite{bekenstein}.

In the spatially homogeneous and isotropic model, the tensor field
equations of motion obtained from Eq. (\ref{eq:motion-g}) are

\begin{equation}
\label{eq:friedmann}
\frac{\dot{a}^2}{a^2}=\frac{8\pi G}{3}\tilde{\rho} e^{-2\phi}+\frac{8\pi }{3k}\mu
\dot{\phi}^2+\frac{2\pi}{3k^2l^2} \mu^2 F(\mu),
\end{equation}

\begin{equation}
\label{eq:acceleration}
2\frac{\ddot{a}}{a}+\frac{{\dot{a}}^2}{a^2}=-8\pi G
\tilde{p}e^{-2\phi}-\frac{8\pi \mu}{k}{\dot{\phi}}^2+\frac{2\pi \mu^2}{k^2 l^2}F(\mu),
\end{equation}

\noindent where an overdot denotes a time derivative. These
equations are the analog of the Friedmann equations in general
relativistic cosmology. The dynamical scalar field equation of
motion obtained from Eq. (\ref{eq:motion-s2}) is

\begin{equation}
\label{eq:scalar-field}
\mu \ddot{\phi}+\Big(3\mu \frac{\dot{a}}{a}+ \dot{\mu}\Big)\dot{\phi}+
\frac{kG}{2}e^{-2\phi}(\tilde{\rho} +3\tilde{p})=0.
\end{equation}

\noindent Additionally, stress-energy conservation obtained from
the Bianchi identity, $\tilde{T}^{{\rm eff}\, \, ;\beta}_{\alpha
\beta}=0$, and Eqs. (\ref{eq:motion-vec})--(\ref{eq:motion-s2})
and (\ref{eq:porque?}) leads to the equation of conservation of
stress-energy of ordinary matter,

\begin{equation}
\label{eq:conservation}
\dot{\tilde{\rho}}=3\Big(\dot{\phi}-\frac{\dot{a}}{a}\Big)(\tilde{\rho} + \tilde{p}).
\end{equation}

\noindent Consistent with the case of general relativistic
cosmology, it may be shown that only three of the four equations
(\ref{eq:friedmann})--(\ref{eq:conservation}) are independent.
Note that $U^{\mu}=\delta^{\mu}_{t}$  causes $U^{[\mu; \nu]}$ to
vanish, so Eqs. (\ref{eq:motion-vec}) and (\ref{eq:porque?}) are
identically satisfied. On the other hand, given an expression for
the free function $F(\mu)$ (which must be externally specified
since there is no theory for its determination), Eq.
(\ref{eq:motion-s1}) determines $\mu$ and $\sigma$ in terms of
$\phi$, $U_{\alpha}$, and $g_{\alpha \beta}$. Equations
(\ref{eq:motion-vec}), (\ref{eq:motion-s1}), and
(\ref{eq:porque?}) may be ignored in the following analysis of the
spatially homogeneous and isotropic  model.

In the following section we look for analytical solutions of the
set of field equations (\ref{eq:friedmann}) and
(\ref{eq:scalar-field}) along with (\ref{eq:conservation}).

\section{solutions}
\label{sec:sol}

\subsection{Vacuum case}
\label{sec:vac}

The vacuum case with $\tilde{\rho}=0=\tilde{p}$ is a simple case
to solve analytically. The Friedmann and scalar field equations
become

\begin{equation}
\label{eq:fried-vac}
\frac{\dot{a}^2}{a^2}= \frac{8 \pi}{3 k}\mu {\dot{\phi}}^2+ \frac{2 \pi}{3 k^2 l^2}\mu^2 F(\mu),
\end{equation}

\begin{equation}
\label{eq:sf-vac}
\mu \ddot{\phi} + \left( 3 \mu \frac{\dot{a}}{a} + \dot{\mu}\right)\dot{\phi}=0.
\end{equation}

\noindent Assuming a time independent $\mu$, we find two exact solutions.

Our first solution corresponds to $a=a_0$ and $\phi(t)= \alpha t +
\phi_0$, where $\alpha = \left[-\mu F/(4kl^2) \right]^{1/2}$.
Since the scale factor is constant, this solution is not very
interesting.

Our second solution is the analog of the general relativistic de
Sitter  solution

\begin{equation}
\label{eq:fe-vac}
a(t) = a_0 e^{\pm H_0 \, t},
\end{equation}

\begin{equation}
\label{eq:sfe-vac}
\phi(t)=\phi_0,
\end{equation}

\noindent where $a_0$ and $\phi_0$ are constants and $H_0=\left[
2\pi \mu^2 F/(3k^2l^2) \right]^{1/2}$. This solution also depends
on the form of the free function $F(\mu)$, which is required to be
positive in this case. This solution corresponds to the ``slow
roll'' approximation solution discussed in Ref. \cite{hao}. It is
interesting that the TeVeS model has an accelerated expansion
solution even in the absence of a cosmological constant or a dark
energy scalar field.

\subsection{Barotropic fluid case}
\label{sec:barotropic}

We assume that the ordinary matter fluid has a barotropic equation
of state  $\tilde{p}=\omega \tilde{\rho}$,  so the stress-energy
conservation equation (\ref{eq:conservation}) has the solution

\begin{equation}
\label{eq:density}
\tilde {\rho}=\frac{e^{3(1+\omega)\phi}}{a^{3(1+\omega)}}.
\end{equation}

In order to find an analytical solution of the field equations, we
make the ansatz

\begin{equation}
\label{eq:scalefactor}
\left[a(t)\right]^{3(1+\omega)} = e^{(1+3\omega)\phi(t)},
\end{equation}

\noindent so $\tilde{\rho}(t)= {\rm exp}(2 \phi)$. The Friedmann
and scalar field equations, (\ref{eq:friedmann}) and
(\ref{eq:scalar-field}), then become

\begin{equation}
\label{eq:f2}
{\dot{\phi}}^2\left[\frac{1}{3}\frac{(1+3\omega)^2}{(1+\omega)^2}-\frac{8\pi}{k}\mu\right]
= 8\pi G + \frac{2\pi}{k^2l^2}\mu^2 F(\mu),
\end{equation}

\begin{equation}
\label{eq:sfe2}
\mu \ddot{\phi}+\frac{1+3\omega}{1+\omega}\mu {\dot{\phi}}^2+\dot{\mu}\dot{\phi}+\frac{kG}{2}(1+3\omega)=0.
\end{equation}

\noindent From Eq. (\ref{eq:f2}) we note that if $\mu$ is time
independent, $\dot{\phi}$ is time independent and so
$\ddot{\phi}=0$. Taking $\ddot{\phi}=0$ considerably simplifies
Eq. (\ref{eq:sfe2}), resulting in an analytical expression for the
scalar field,

\begin{equation}
\label{eq:scalarfield}
\phi(t)= \pm \, \alpha \, t  \,  + \, \phi_0,
\end{equation}

\noindent where $\phi_0$ is a constant of integration and now
$\alpha=\left[-kG(1+\omega)/(2\mu)  \right]^{1/2}$. Using this
solution for $\phi(t)$ in the Friedmann equation (\ref{eq:f2}) we
get a relation between the constants $\omega$ and $\mu$ and the
free function $F(\mu)$,

\begin{equation}
\label{eq:f-w}
\frac{2\pi}{k^2l^2}{\mu}^2 F(\mu)=-\frac{kG(1+3\omega)^2}{6\mu(1+\omega)}-4\pi G(1-\omega).
\end{equation}

\noindent It is important to note that this equation does not give
us a function $F(\mu)$, but instead given an $F(\mu)$ it relates
$\mu$ and $\omega$, or given an $F(\mu)$ and an $\omega$ it fixes
the  value of $\mu$.

From Eqs. (\ref{eq:scalefactor}) and (\ref{eq:scalarfield}) we see
that our solution is a de Sitter solution with scale factor

\begin{equation}
\label{eq:sf}
a(t)=a_{0} e^{\pm H_0 t},
\end{equation}

\noindent where

\begin{equation}
\label{eq:H}
H_0 = \sqrt{-\frac{kG(1+3\omega)^2}{18\mu(1+\omega)}},
\end{equation}

\begin{equation}
\label{eq:a0}
a_{0} = e^{\frac{1+3 \omega}{3(1+\omega)} \phi_0 }.
\end{equation}

\noindent Recalling the definition $\mu= k G \sigma^2$, we get
$H_0 = (1+3\omega)/\left[-18\sigma^2 (1+\omega)\right]^{1/2}$.
Since $H_0$ must be real, our solution must have $\omega < -1$.
This is not necessarily a problem, since we are interested in a
solution that may be applicable during inflation or the current
accelerated expansion epoch. It is interesting that
Eq.(\ref{eq:sf}) is a de Sitter solution, independent of the value
of $\omega$, provided $\omega < -1$.

Using this solution in the field equations, we find the following useful identities,

\begin{equation}
\label{eq:einst-1b}
H_0^2=\frac{8\pi G}{3}\left[1 + \frac{\mu
\alpha^2}{kG}+\frac{\mu^2}{4l^2k^2G}F(\mu)\right],
\end{equation}

\begin{equation}
\label{eq:scalar-b}
3H_0\alpha+\frac{kG}{2\mu}(1+3\omega)=0.
\end{equation}

The explicit contribution of the scalar field to the energy
density in this model is not easily seen because ordinary matter
and the scalar field are entangled, not only in the metric, Eq.
(\ref{eq:phys-met}), but also in the right hand side of Eq.
(\ref{eq:motion-g}). $\tilde{T}_{00}=\tilde{\rho}\,e^{2\phi}$ is
the energy density of ordinary matter so  $\tilde{T}^{{\rm
eff}}_{00}-\tilde{T}_{00}=-\tilde{\rho}e^{2\phi}+\tilde{\rho}e^{-2\phi}+\mu
\dot{\phi}^2/(kG)+ \mu^2 F/(4k^2 l^2 G)$ could be associated with
the contribution of the scalar field to the energy density, as in
Brans-Dicke-like models \cite{wagoner}, but note that in
$\tilde{T}^{{\rm eff}}_{00}-\tilde{T}_{00}$ the stress-energy
tensor of ordinary matter also contributes. Of course,
$\tilde{T}^{{\rm eff}}_{00}$ is the right hand side of the
Friedmann equation (\ref{eq:friedmann}), which we rewrite as

\begin{equation}
\label{eq:friedmann2}
\frac{\dot{a}^2}{a^2}=\frac{8\pi G}{3}\rho_{\rm eff},
\end{equation}

\noindent and similarly the spatial components $\tilde{T}^{\rm
eff}_{ij}$ on the right hand side of Eq. (\ref{eq:acceleration})
may be rewritten as

\begin{equation}
\label{eq:acceleration2}
\frac{\ddot{a}}{a}=-\frac{4\pi G}{3}\left[\rho_{\rm eff}+ 3 p_{\rm eff}\right],
\end{equation}

\noindent where

\begin{equation}
\label{eq:rho-eff}
\rho_{\rm eff}=\tilde{\rho} e^{-2\phi} + \frac{\mu}{kG}{\dot{\phi}}^2+\frac{\mu^2}{4k^2l^2G} F(\mu),
\end{equation}

\begin{equation}
\label{eq:p-eff}
p_{\rm eff}=\tilde{p} e^{-2\phi} + \frac{\mu}{kG}{\dot{\phi}}^2-\frac{\mu^2}{4k^2l^2G} F(\mu).
\end{equation}

\noindent $\rho_{\rm eff}$ and $p_{\rm eff}$ obey the energy
conservation equation $\dot{\rho}_{\rm eff}=-3(\rho_{\rm
eff}+p_{\rm eff})\dot{a}/a$. It may be shown that when $\rho_{\rm
eff}$ and $p_{\rm eff}$, Eqs. (\ref{eq:rho-eff}) and
(\ref{eq:p-eff}), are used in this spatially homogeneous
conservation equation, we recover the spatially homogeneous part
of the scalar field equation of motion, Eq.
(\ref{eq:scalar-field}). See, {\it e.g.}, Ref. {\cite{ratra2}} for
the general relativistic cosmology analog of this result.

Note that in Ref. \cite{hao} $\rho_m= \tilde{\rho}e^{-2\phi}$ is
defined to be the energy density of ordinary matter, and
$\rho_{\phi}=\mu \dot{\phi}^2/(kG)+\mu^2 F/(4k^2 l^2 G)$  the
effective energy density of the scalar field $\phi$. However, we
have shown here that the energy density of ordinary matter is
given by the time-time component of the stress-energy tensor of
ordinary matter $\rho_m=\tilde{T}_{00}=\tilde{\rho}\,e^{2\phi}$.
The definition of $\rho_{\phi}$ in Ref. \cite{hao} corresponds to
a piece of  $\, \tilde{T}^{{\rm eff}}_{00}-\tilde{T}_{00}$, that
may be associated with the energy density of the scalar field.
Thus, the energy densities $\rho_m$ and $\rho_{\phi}$ defined in
\cite{hao} do not correspond to the energy densities of a fluid in
the sense that they do not obey an energy conservation equation
obtained from the Bianchi identity, which reduces to an equation
of motion for the scalar field. These definitions are however, a
piece of the effective energy density defined above.

Our solution, Eqs. (\ref{eq:scalarfield})--(\ref{eq:sf}), implies
that $\rho_{\rm eff}=-k(1+3\omega)^2/[48\pi \mu(1+\omega)]$ is
positive for $\omega < -1$.

We will  examine the stability of our solution in more detail in the following section.

\subsection{Bekenstein's $F(\mu)$ function}
\label{sec:F-fuction}

The exact solutions that we have obtained are valid for any
$F(\mu)$ function, as long as the function satisfies the relation
given in Eq. (\ref{eq:f-w}) for the solution corresponding to a
barotropic equation of state, Sec. \ref{sec:barotropic}, and
$F(\mu) > 0$ for the solution corresponding to the vacuum case,
Sec. \ref{sec:vac}. As pointed out by Bekenstein,
\cite{bekenstein}, there is no theory for the function $F(\mu)$,
and based on this freedom and other considerations he picks as an
example,

\begin{equation}
\label{f-bekenstein}
F( \mu)=\frac{3}{8\mu}(4+2 \mu-4 \mu^2 + \mu^3)+\frac{3}{4\mu^2}\ln[(1-\mu)^2].
\end{equation}

\noindent The cosmologically relevant part of this function lies
in the interval $ 2 < \mu < \infty$. Using this expression, from
Eq. (\ref{eq:f-w}) we confirm that our solution corresponding to a
barotropic perfect fluid is limited to $\omega \leq -1$. We also
find that $F(\mu)>0$ for any value of $\mu$ in the interval
relevant to cosmology, so this form of $F(\mu)$ can also be used
in the solution for the vacuum case.

\subsection{Density and deceleration parameters}
\label{sec:omega-q}

We rewrite the Friedmann equation, Eq. (\ref{eq:friedmann}), as

\begin{equation}
\label{eq:omega}
1=\Omega_m+\Omega_{\phi},
\end{equation}

\noindent where the density parameters of ordinary matter and the scalar field are,

\begin{eqnarray}
\label{eq:omega-m}
\Omega_m &=& \frac{8\pi G}{3H^2}\tilde{\rho}\, e^{2\phi}, \\
\label{eq:omega-phi}
\Omega_{\phi} &=& \frac{8\pi G}{3H^2}\Big[-\tilde{\rho}\, e^{2\phi}+\tilde{\rho}\,
e^{-2\phi} + \frac{\mu}{kG}\dot{\phi}^2+\frac{\mu^2}{4k^2l^2G}F(\mu)\Big],
\end{eqnarray}

\noindent where $H=\dot{a}/a=\tilde{H} + \dot{\phi}\, e^{-\phi}$, 
$\tilde{H}=(d\tilde{a}/d\tilde{t})/ \tilde{a}$. As in general relativistic 
cosmology, we use Eqs. (\ref{eq:friedmann}) and
(\ref{eq:acceleration}) to determine the deceleration parameter,
$q= -\ddot{a}/a H^2$, finding

\begin{equation}
\label{eq:q}
q=\frac{4\pi G}{3H^2}\tilde{\rho}\, e^{-2\phi}(1+3\omega)+\frac{8\pi G}{3 H^2 }\left[\frac{2\mu}{kG}\dot{\phi}^2-
\frac{\mu^2}{4k^2l^2G}F(\mu)\right].
\end{equation}

\noindent For our barotropic perfect fluid solution, Eqs.
(\ref{eq:scalarfield})--(\ref{eq:a0}), Eq. (\ref{eq:q}) gives
$q=-1$ independent of the form of the $F(\mu)$ function and the
value of $\omega$. The same result also holds for our solution in
the vacuum case (Sec. \ref{sec:vac}). Of course, both results
follow directly from Eqs. (\ref{eq:fe-vac}) and (\ref{eq:sf}) and
the definition of $q$.

\subsection{Physical variables}
\label{sec:physicalframe}

Until this point we have been working in the Einstein metric, {\it
i.e.}, in the metric $g_{\mu\nu}$  whose dynamics is governed by
the Einstein-Hilbert action. We will call this frame the
``Einstein frame''. It is important to write the solutions
found above in terms of the physical variables, {\it i.e.},
in the frame of the metric $\tilde{g}_{\mu\nu}$  known as
``physical frame'' or what
Skordis {\it et al.} \cite{skordis} called the ``matter frame''. These metrics are related by
Eq. (\ref{eq:phys-met}), so for the spatially-flat
Friedmann-Lema{{\^\i}}tre-Robertson-Walker metric of Eq. (\ref{eq:metric})
 we have $\tilde{a}=e^{-\phi}a$ and
$\tilde {dt}=e^{\phi}dt$. In the ``physical frame'' we define
the deceleration parameter in terms of physical variables as

\begin{equation}
\label{eq:q-pf} q=-\frac{\tilde{a}(\tilde{t})
\ddot{\tilde{a}}(\tilde{t})}{\left(\dot{\tilde{a}}(\tilde{t})\right)^2}
\end{equation}

\noindent where now an overdot denotes differentiation
with respect to physical time $\tilde{t}$.

\begin{itemize}

\item{Vacuum case.}

The solution of the vacuum case in the Einstein frame,  Eqs. (\ref{eq:fe-vac})
and (\ref{eq:sfe-vac}), corresponds to the physical frame solution

\begin{equation}
\label{eq:vac-pf}
\tilde{a}(\tilde{t})=\tilde{a}_0 e^{\pm \tilde{H}_0 \tilde{t}},
\end{equation}

\noindent where $\tilde{a}_0=a_0 e^{-\phi_0}$ and $\tilde{H}_0=H_0 e^{-\phi_0}$.
The vacuum solution in the physical variables is also a de Sitter solution
with $q=-1$, as can be seen from Eq. (\ref{eq:q-pf}).

\item{Barotropic ideal fluid case.}

We have found that the solution for a barotropic perfect fluid in the Einstein
frame is $a(t)=a_0 e^{\pm H_0 t}$ and $\phi(t)= \pm \, \alpha t + \phi_0$, where $H_0$ and
$\alpha$ are given by Eqs. (\ref{eq:H}) and (\ref{eq:scalarfield}). Rewriting
this solution in terms of the physical variables we get

\begin{equation}
\label{eq:baro-pf}
\tilde{a}(\tilde{t})=\tilde{a}_0 \, \tilde{t}^{H_0/\alpha -1}=
\tilde{a}_0 \, \tilde{t}^{- 2/3(1+\omega)},
\end{equation}

\begin{equation}
\label{eq:scalar-pf} \phi(\tilde{t})=\ln(\alpha\tilde{t})
\end{equation}

\noindent where $\tilde{a}_0=a_0 \, e^{-\phi_0 H_0/\alpha}(
\alpha)^{ (H_0/\alpha -1)}= [-kG(1+\omega)/(2\mu)]^{-
1/3(1+\omega)}$. Note that the contracting solution in the Einstein frame does 
not have physical meaning in the physical variables. As we can see, the de Sitter 
solution in the Einstein frame for a barotropic ideal fluid is transformed to 
a power law solution in the physical frame. Since  $\omega < -1$ the solution is
expanding.

Using Eq. (\ref{eq:q-pf}) it is straightforward to check that the
deceleration parameter in the physical frame for the solution (\ref{eq:baro-pf}) 
is $q=-(5+3\omega)/2$, then $q < 0$ implies $\omega > -5/3$. Since $\omega < -1$ this solution 
expands in an accelerated way for $\omega$ in the interval of values $-5/3 < \omega < -1$. 
Such solutions have been discussed in the context of power law inflation \cite{abbott}.

\end{itemize}

\section{stability of the solutions}
\label{sec:stability}

In Sec. \ref{sec:sol} we have found exact solutions for the vacuum
case, Eqs. (\ref{eq:fe-vac}) and (\ref{eq:sfe-vac}), and for the
case with a barotropic ideal fluid, Eqs. (\ref{eq:scalarfield})
and (\ref{eq:sf}). The purpose of this section is to see if these
solutions are stable with respect to small perturbations.

A complete analysis will require solving the TeVeS cosmological
linear perturbation equations (see Refs. \cite{skordis, skordis2})
for small departures from the spatially homogeneous and isotropic
background solutions that we have derived in the previous section.
In our preliminary exploration here we will focus only on very
large scale perturbations, using instead the method developed in
Ref. \cite{ratra}.

We first consider the barotropic ideal fluid case. Following the
method of Ref. \cite{ratra} we make the change of variables,

\begin{equation}  \label{eq:pert}
\begin{aligned} a(t)&=a_e(t)v(t),\\ \phi(t)&=\phi_e(t)+\psi(t),\\
\tilde{\rho}(t)&=\tilde{\rho}_e(t)+\eta(t). \end{aligned}
\end{equation}

\noindent Here $v(t)-1$, $\psi(t)$, and $\eta(t)$ are small
perturbations about the spatially homogeneous solution, denoted
here as $a_e$, $\phi_e$, and $\rho_e$, and given in Eqs.
(\ref{eq:sf}), (\ref{eq:scalarfield}), and (\ref{eq:density}).
Defining two ``speeds'' $s=\dot{\psi}$ and $r=\dot{v}$ (note that
$\tilde{\rho}$ obeys a first order differential equation and so we
do not need a ``speed'' for it), the equations of motion become,

\begin{equation}
\label{eq:pert-1}
\begin{aligned}
\dot{\psi}&=s,\\
\dot{s}&=-\frac{3\alpha}{v\mu}r-\frac{\dot{\mu}}{\mu}\alpha-\left(3H_0 +3 \frac{r}{v}+
\frac{\dot{\mu}}{\mu}\right)s-\frac{kG(1+3\omega)}{2\mu} e^{-2\phi_e}
\left(e^{-2\psi}(\tilde{\rho}_e+\eta)-\tilde{\rho}_e\right), \\
\dot{v}&=r, \\
\dot{r}&=-2H_0r-\frac{4\pi G}{3}(1+3\omega)e^{-2\phi_e}\left(e^{-2\psi}(\tilde{\rho}_e+\eta)-
\tilde{\rho}_e\right)v-\frac{4\pi G}{3}v \left( \mathcal{F}(\alpha+s)-\mathcal{F}(\alpha) \right), \\
\dot{\eta}&=-3\left(\frac{r}{v}-s\right)(1+\omega)\tilde{\rho}_e-3\Big(H_0+\frac{r}{v}-\alpha-s\Big)\eta,
\end{aligned}
\end{equation}

\noindent where $\mathcal{F}= \mu \dot{\phi}^2/(kG)+ \mu^2 F(\mu)/(4l^2 k^2 G)$.

The vector field $U^\alpha$ contributes to the initial set of
equations (\ref{eq:motion-vec})--(\ref{eq:porque?}), but it does
not contribute to the spatially homogeneous cosmological equations
(\ref{eq:friedmann})--(\ref{eq:conservation}), because there is no
preferred spatial direction on cosmological scales and so
$U^\alpha=\delta^\alpha_t$. To simplify the analysis we do not
consider a perturbation of $U^\alpha$ here.

$\mu(\dot{\phi})$ is a function of the scalar field's time derivative and is time 
independent for our background solutions $\mu=\mu(\alpha)$. When we perturb the scalar field, 
$\mu=\mu(\alpha+s)$ becomes time dependent. Since we are interested in linear perturbation 
theory, we expand it in a Taylor series with respect to the small parameter $s$,
and we keep only terms up to first order $\mu(\alpha+s)=\mu(\alpha)+ s\, d\mu/{d\alpha}$.

The critical point of Eqs. (\ref{eq:pert-1}) is
$(\psi_0,s_0,v_0,r_0,\eta_0)\, = \, (0,0,\bar{v},0,0)$, where
$\bar{v}$ is an arbitrary constant which corresponds to the
freedom in rescaling $a_0$. This point of phase space is where all
time derivatives in Eqs. (\ref{eq:pert-1}) vanish. Perturbing Eqs.
(\ref{eq:pert-1}) around the critical point,
$(\psi,s,v,r,\eta)=(\psi_0+\psi_1,s_0+s_1,v_0+v_1,r_0+r_1,
\eta_0+\eta_1)$, the linear perturbation equations are,

\begin{equation}  \label{eq:pert-2}
\begin{aligned}
\dot{\psi_1}&=s_1,\\
\dot{s}_1&=\frac{kG}{\mu_0}(1+3 \omega)\psi_1-3H_0s_1-
\frac{3\alpha}{\bar{v}\mu_0}r_1-\frac{kG}{2\mu_0}(1+3 \omega)e^{-2\phi_e}\eta_1, \\
\dot{v_1}&=r_1, \\
\dot{r}_1&=\frac{8\pi G}{3}(1+3\omega)\bar{v}\psi_1- \frac{4\pi
G}{3}\frac{d\mathcal{F}}{d\alpha}\bar{v}s_1-2H_0r_1
-\frac{4\pi G}{3}(1+3\omega)e^{-2\phi_e}\bar{v}\eta_1, \\
\dot{\eta}_1&=3(1+\omega)\tilde{\rho}_es_1-3\frac{(1+\omega)\tilde{\rho}_e}
{\bar{v}}r_1-3(H_0-\alpha)(1+\omega) \eta_1,
\end{aligned}
\end{equation}
 
\noindent where $\mu_0=\mu(\alpha)$.

In order to compute the eigenvalues of the matrix defined by these
equations, we need to specify the function $F(\mu)$ and the values
of the constants of the TeVeS model. As an example, we consider
Bekenstein's \cite{bekenstein} $F(\mu)$ function, Eq.
(\ref{f-bekenstein}), for some values of $\mu$ in the cosmological
regime $2 < \mu < \infty$. We choose two set of arbitrary values
for the parameters of the model, $k=l=G=1$ and $k=0.1, \, l=10, \,
G=1$. Also, we consider three illustrative values for
$\omega=-1.02, \, -1.05$, and $-1.08$. Numerical computation shows
that the expanding solution, $a(t)=a_0 e^{H_0\, t}$, is stable for
small values of $\mu$ ($\mu < 19$ when $\omega=-1.02$, $\mu < 13$
when $\omega=-1.05$, and $\mu < 11$ when $\omega=-1.08$) for
$k=l=G=1$. For the second set of parameters $k=0.1, \, l=10, \,
G=1$ we find that the expanding solution is stable for small
values of $\mu$ ($\mu < 11$) for all three values of $\omega$ that
we consider. The contracting solution, $a(t)=a_0 e^{-H_0 \, t}$,
on the other hand, is not stable for any $\mu$ in the cosmological
allowed range, independent of the values of the other model
parameters we consider. In the limit of small deviation from
$\omega=-1$, analytical computations show that the expanding
solution $a(t)=a_0 e^{H_0\, t}$ is stable. Clearly, the stability
of the solution depends sensitively on the form of $F(\mu)$ and
the values of $k, \, l, \, G$, and $\mu$. We note that even though
$\omega < -1$, this does not necessarily imply instability.

The expanding solution for the vacuum case,  $a(t)=a_0 e^{H_0 \,
t}$ with $\phi(t)=\phi_0$, is stable with linear perturbation
eigenvalues $\lambda_1=-2H_0$, $\lambda_2=-3H_0$. The contracting
solution, $a(t)=a_0 e^{-H_0 \, t}$, is unstable with eigenvalues
$\lambda_1=2H_0$, $\lambda_2=3H_0$. These results hold for any
$F(\mu)>0$ function.

We have calculated analytically the eigenvalues of small
perturbations in the limit of general relativity. We find that for
both the vacuum and barotropic equation of state cases, the
expanding solutions $a(t)= a_0 e^{H_0\, t}$ are stable, as well
known.

\section{Conclusions}
\label{sec:conclusion}

We have found two homogeneous and isotropic cosmological solutions
of the TeVeS model, a relativistic generalization of MOND. In the vacuum case, 
the analytical solutions in both, Einstein and physical frames, are de Sitter 
solutions with $q=-1$, and the scalar field is a constant $\phi(t)=\phi_0$. 
For a barotropic ideal fluid we find a de Sitter solution in the Einstein frame, 
which is transformed to a power law expanding solution in the physical frame. 
Accelerated expansion for this solution occurs for $-5/3 < \omega < -1$. 
The scalar field in this case evolves logarithmically with the physical time, 
$\phi(\tilde{t})=\ln(\alpha\tilde{t})$.

The free function $F(\mu)$ plays an important role in the TeVeS
model. However, we have found that for vacuum case the only
restriction is $F(\mu)>0$, while for the barotropic perfect fluid
case our solutions are valid for a family of $F(\mu)$ functions
that obey the $\omega$ dependent relationship given in Eq. (\ref{eq:f-w}).

In the vacuum case the expanding solution $a(t)=a_0 e^{Ht}$ is
stable for any $F(\mu)>0$ function. In the more general case of a
barotropic ideal fluid, we have studied stability for a few values
of the parameters and constants of the theory and have assumed
Bekenstein's $F(\mu)$ function. Typically, the expanding solution,
$a=a_0 e^{Ht}$ in Einstein frame, is stable for small values of
$\mu$, and becomes unstable for large values of $\mu$. Since the
solutions in Einstein frame and physical frame are related
through transformation of variables, the corresponding solution
in the physical frame $\tilde{a}(\tilde{t})=\tilde{a}_0 \,
\tilde{t}^{H_0/\alpha -1}$ has the same stability properties.

The accelerated expansion TeVeS cosmological solutions we have
found might prove helpful when trying to incorporate inflation or
the present epoch of accelerated expansion in the TeVeS scenario.

\begin{acknowledgments}

We are indebted to the referee for very useful comments which helped us 
to clarify and improve our results. We acknowledge support from DOE EPSCoR 
grant DE-FG02-00ER45824 and DOE grant DE-FG03-99ER41093.

\end{acknowledgments}


\end{document}